\documentclass[10pt]{ws-procs9x6}

\begin{document}

\title{Resummation in hot field theories}

\author{U. Reinosa}

\address{Laboratoire de Physique Th\'eorique, \\
CEA/Saclay, Orme des Merisiers, \\ 
91191 Gif sur Yvette, FRANCE\\ 
E-mail: reinosa@spht.saclay.cea.fr}

\maketitle

\abstracts{We consider a scalar theory at finite temperature in the 2PI resummation scheme, including $\phi^3$ and $\phi^4$ interactions. Already at the one loop level in this scheme, we have to deal with a non local approximation. We carry out the renormalization and obtain finite equations for the propagator. Within this model we can explore the effect of non local contributions to the self-energy in the evaluation of thermodynamic quantities.}

\section{Introduction}
Self-consistent approximation schemes based on two particle irreducible functionals have proved to be successful for studying systems where the quasi-particle description gives a good understanding of the relevant degrees of freedom. This is for example the case for thermodynamic properties of the quark-gluon plasma [1]. These resummations schemes have also been recently applied to study the dynamics of quantum fields out of equilibrium [2].

In these schemes, any physical quantity is expressed in terms of the full propagator and the approximations used to compute the physical quantity and the propagator are ``self-consistently'' correlated with each other. For instance, for applications to thermodynamics, one starts by expressing the thermodynamic potential in terms of the full propagator [3,4,5]. A central quantity is then the sum of the 2 PI skeleton diagrams: $\Phi[D]$. The full self-energy $D$ is then given by:
\begin{equation}\label{eq:gap}
\Pi=2\frac{\delta \Phi}{\delta D},
\end{equation}
which, together with Dyson's equation, defines the full propagator in a self-consistent manner.

One of the main questions to be addressed before any numerical computation is that of renormalizability. Via the self-consistent equation for the self-energy, one is effectively resumming an infinite set of diagrams together with their UV singularities. One has then to devise a procedure where all the UV singularities disappear at once. The problem becomes even worse at finite temperature. In fact, since the UV singularities depend on the self-energy itself, when turning on temperature they become $T$-dependent.

A major progress was recently achieved by H. van Hees and J. Knoll [6,7], who showed, in the real time formalism, how to renormalize in such a way that temperature dependent counterterms never appear. In a recent work [8], we have extended the analysis of van Hees and Knoll using the imaginary time formalism, and clarifying some aspects of renormalization of the vacuum sector giving an explicit construction of counterterms. In this short communication, we restrict ourselves to a simple model where only parts of the difficulties dealt with in [6,7,8] appear.

\section{One loop approximation in a scalar toy model}
\begin{figure}
\centerline{\epsfxsize=4in\epsfbox{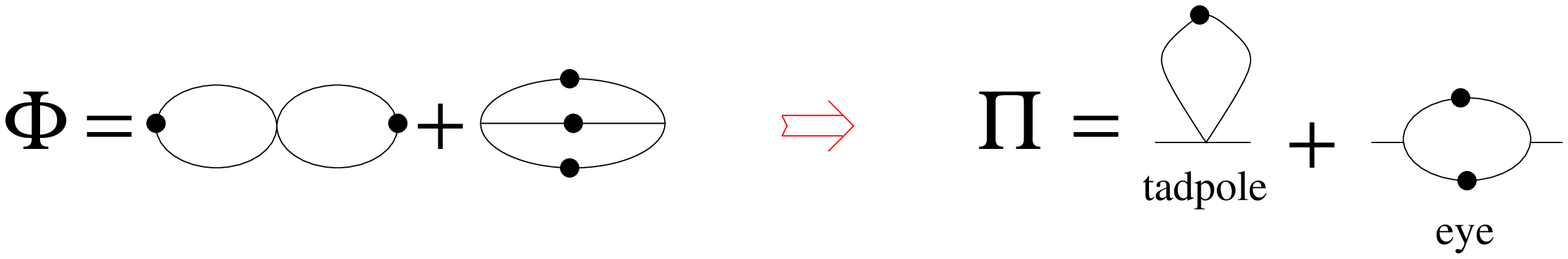}}
\caption{Two loop approximation for $\Phi$ and related gap equation for $\Pi$}
\label{fig:approx}
\end{figure}

We work in a scalar theory with $\phi^3$ and $\phi^4$ interactions in $d=4$ space-time dimensions. To two loop order, the $\Phi$ functional is given in figure \ref{fig:approx}.
The gap equation is obtained by cutting one line on each diagram. We then obtain a one loop approximation for the self-energy. The diagrams entering this equation are given by the following sum-integrals:
\begin{equation}
\Pi_{tad}=\frac{\lambda_0}{2}\int_{\omega_n,p}\hspace{-0.8cm}\Sigma D(i\omega_n,p),
\end{equation}
for the tadpole diagram, and
\begin{equation}
\Pi_{eye}(i\omega,k)=-\frac{g^2}{2}\int_{\omega_n,p}\hspace{-0.8cm}\Sigma D(i\omega_n,p)D(i\omega-i\omega_n,k-p),
\end{equation}
for the eye  diagram. In four dimensions, these integrals are divergent. One expects the effects of the $\phi^3$ interactions (here $\Pi_{eye}$) to be super-renormalizable, while those of the $\phi^4$ interactions ($\Pi_{tad}$) to be just renormalizable. The ultraviolet singularities should be absorbed in the usual counterterms $\delta Z$, $\delta m^2$ and $\delta \lambda$. Those have to be defined in the vacuum and be compatible with the underlying perturbative diagrams that are resummed by the equation.

\subsection{Large momentum behaviour of $\Pi(k)$}
To understand the UV structure of the gap equation, one first studies the large momentum behavior of the self energy: 
\begin{itemize}
\item the contribution from the tadpole is a T-dependent mass: $m_{\infty}^2$;
\item the eye contribution consists in a vacuum piece which is logarithmic and a finite temperature piece which behaves like $\frac{T^2}{k^2}$. This last piece plays no role in the analysis of UV singularities and can be ignored in what follows;
\end{itemize}
\begin{equation}
\Pi(k) \sim m_{\infty}^2(T)+m^2\log k^2 \mbox{ , for large k.}
\end{equation}
In particular, there are no corrections in $k^2$ to $\Pi$, so that $\delta Z=0$.

\subsection{Analysis of divergences}
We first split each integral into a vacuum piece and a thermal piece:
\begin{eqnarray}
\Pi_{tad} & = & \frac{\lambda_0}{2}\int_{T=0}(dp)\frac{1}{p^2+m^2+\Pi}+\Pi^T_{tad},\\
\Pi_{eye} & = & -\frac{g^2}{2}\int_{T=0}(dp)\frac{1}{(p^2+m^2+\Pi)((k-p)^2+m^2+\Pi)}+\Pi^T_{eye}\nonumber.
\end{eqnarray}
$\Pi^T_{tad}$ and $\Pi^T_{eye}$ are thermal 1 loop integrals and so are finite in the UV. The explicit integrals are divergent and depend on the temperature via the self-energy.

The eye diagram is logarithmically divergent, but the dominant behavior ot the integrand does not depend on $\Pi$. Therefore, the divergence is independent of T, and can be removed by the standard one-loop counter-term of the vacuum perturbation theory:
\begin{equation}
\delta m_{eye}^2=-\frac{g^2}{32\pi^2}\frac{1}{\varepsilon}.
\end{equation}\vspace{0.2cm}

\begin{figure}
\centerline{\epsfxsize=0.5in\epsfbox{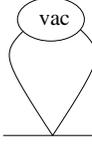}}
\caption{Insertion of the eye diagram into the tadpole, divergent contribution}
\label{fig:eye}
\end{figure}

The tadpole diagram is quadratically divergent and thus sensitive to the subleading behavior of the propagator, that is, to the self-energy $\Pi$. This is the origin of the $T$-dependent singularities. We can in fact be more precise: $T$-dependent singularities only arise from the insertion of the tadpole itself. This is because the insertion ot the dominant piece of the eye diagram ($\Pi_{eye}^{(0)}$) (see figure \ref{fig:eye})
leads to a $T$-independent singularity, i.e., a new contribution to the mass counterterm. The equation for the tadpole can then be written as follows:
\begin{eqnarray}\label{eq:gap}
\Pi_{tad} & = & \lambda_0\int_{T=0}(dp)\left(\frac{1}{p^2+m^2}-\frac{\Pi_{tad}+\Pi_{eye}^{(0)}}{(p^2+m^2)^2}\right)\nonumber\\\nonumber\\
& + & \lambda_0 F(\Pi)+\delta m^2_{tad},
\end{eqnarray}
where $F(\Pi)$ is a finite functional of $\Pi$. $\delta m_{tad}^2$ absorbs usual mass singularities and the new one arising from the eye insertion:
\begin{equation}\label{eq:ct}
\delta m^2_{tad}=\lambda_0\int_{T=0}(dp)\left(\frac{1}{p^2+m^2}-\frac{1}{p^2+m^2}\Pi_{eye}^{(0)}\frac{1}{p^2+m^2}\right).
\end{equation}
The remaining terms need much more attention as they carry the $T$-dependent singularities. We know from the $g=0$ case that the divergence proportionnal to $\Pi_{tad}$ is absorbed in the bare coupling constant. In order to renormalize the ensuing equation, namely (cf. eqs (\ref{eq:gap}) and (\ref{eq:ct})):
\begin{equation}
\Pi_{tad}\left\{\frac{1}{\lambda_0}+\int_{T=0}(dp)\frac{1}{\left(p^2+m^2\right)^2}\right\}=F(\Pi),
\end{equation}
we set
\begin{equation}
\frac{1}{\lambda}=\frac{1}{\lambda_0}+\int_{T=0}(dp)\frac{1}{\left(p^2+m^2\right)^2},
\end{equation}
which is $T$-independent.

Thus we have managed to renormalize the gap equation by avoiding $T$-dependent counterterms. The renormalized gap equations read:
\begin{eqnarray}
\tilde{\Pi}_{eye} & = & \Pi_{eye}+\delta m^2_{eye},\\
\Pi_{tad} & = & g^2F(\Pi_{tad}+\tilde{\Pi}_{eye}).
\end{eqnarray}

\section{First numerical computations}
\begin{figure}
\begin{minipage}{0.49\linewidth}
\begin{center}
\centerline{\epsfxsize=1.6in\epsfbox{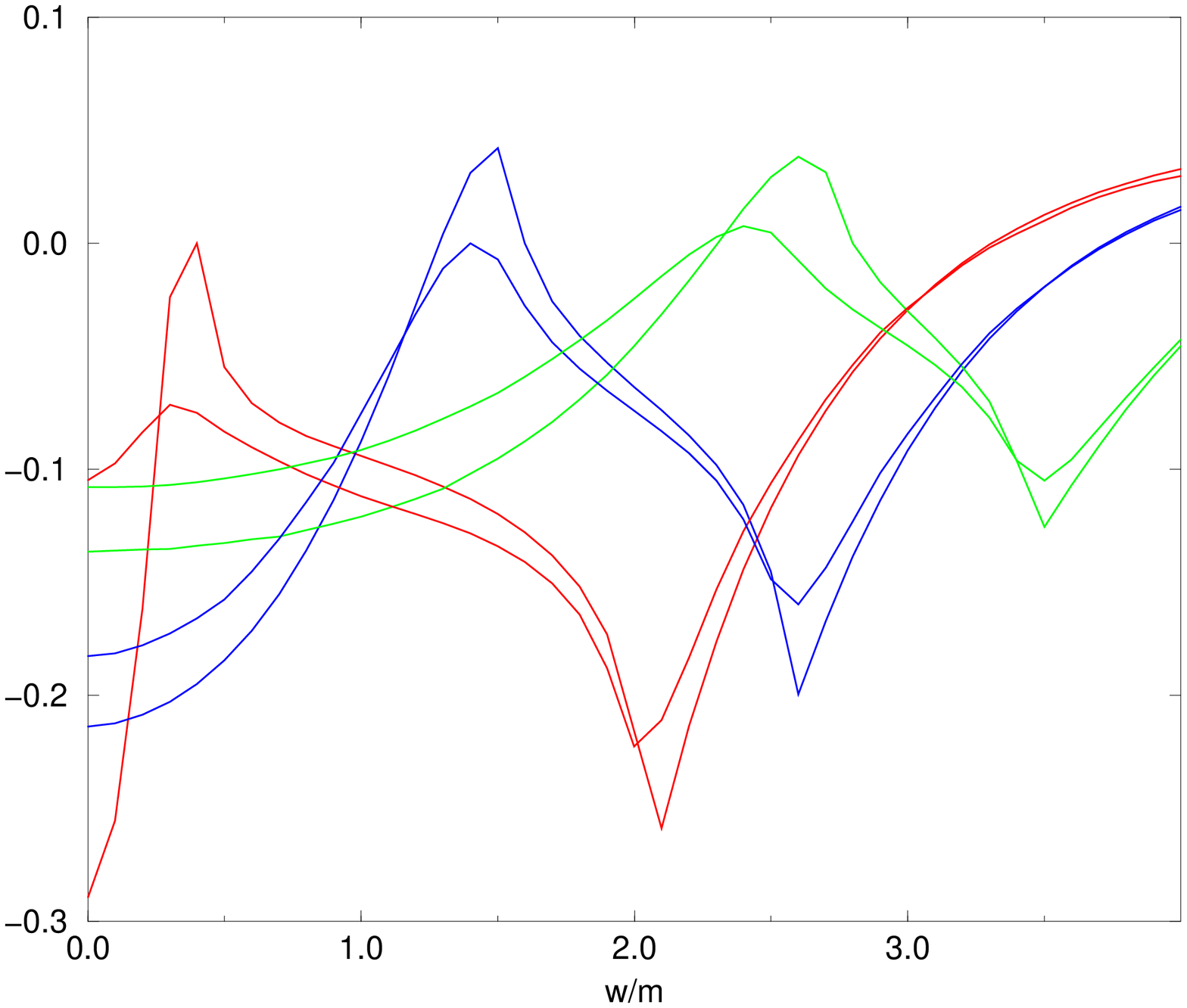}}
\end{center}
\end{minipage} 
\begin{minipage}{0.49\linewidth}
\begin{center}
\centerline{\epsfxsize=1.6in\epsfbox{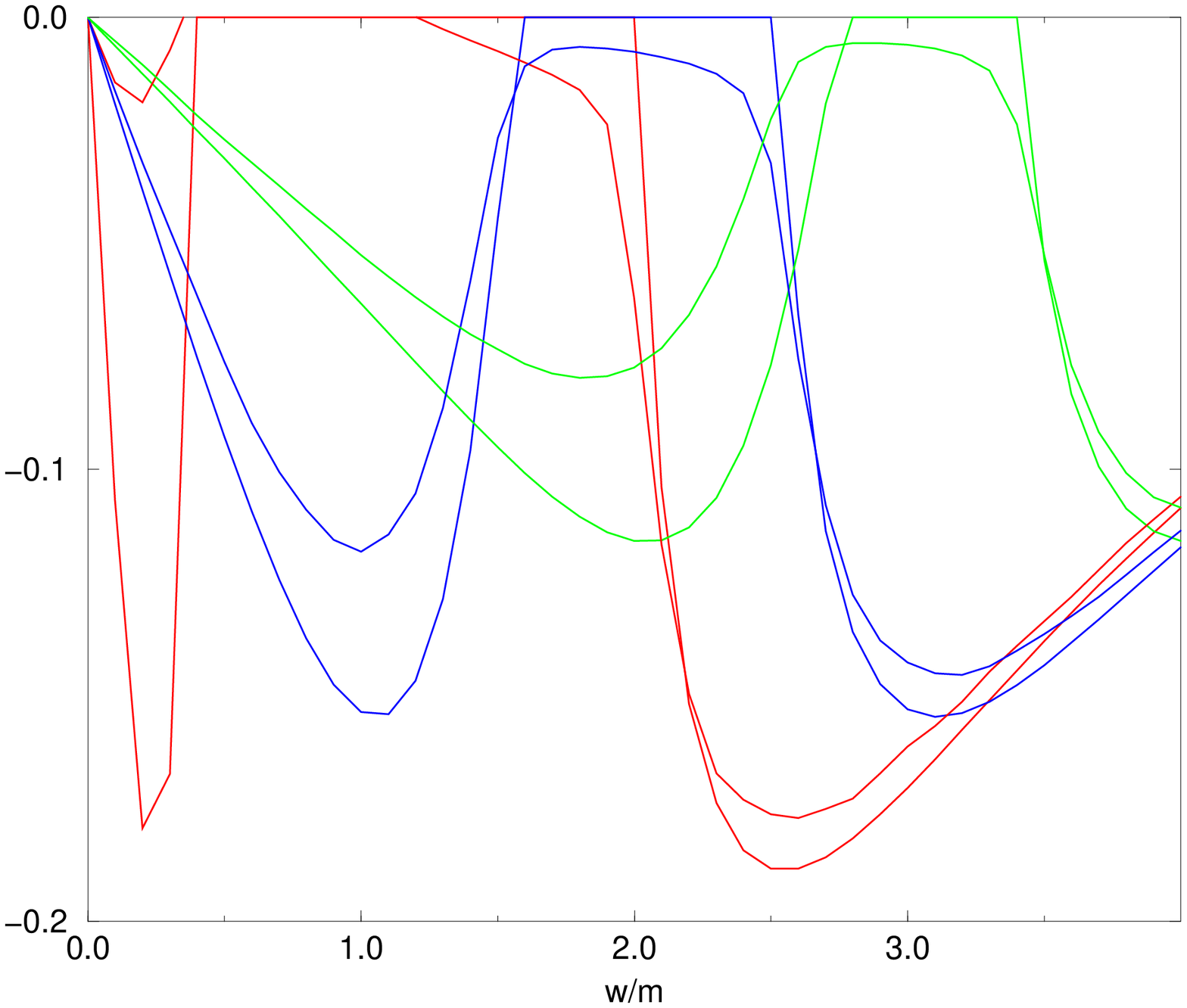}}
\end{center}
\end{minipage} 
\label{fig:iter}
\caption{Resummed vs perturbative self energy. The self energy is plotted as a function of $\omega$ for 3 diferent values of k. The plots go always by 2: perturbative versus resumed result. The first figure represents the real part and the second the imaginary part of $\Pi$.}
\end{figure}

With this finite set of equations, we can evaluate the full self-energy in this particular approximation and compare it to perturbative results. This is shown in figure \ref{fig:iter}. Note that, for this particular model, the difference between the perturbative and the self-consistent results remains rather small because the well known infrared instability induced by $\phi^3$ interactions forces us to restrict ourselves to a very limited range of parameters. But the results in figure \ref{fig:iter} demonstrate that, with our method the non local numerical problem is fully under control, and encourage us to use this method to address similar problems which are more directly relevant for physic, like QCD. New questions will arise, related to gauge invariance (see for example [9]).
\vspace{-0.7cm}
\vspace{-1.0cm}
\section{Conclusion}

We have shown in a simple model how to perform renormalization of the self-consistent approximation in the presence of non-local effects.

\end{document}